\documentclass{elsart}

\usepackage{graphicx}
\usepackage{amssymb}
\usepackage[latin1]{inputenc}
\usepackage{hyperref}

\begin{document}

\begin{frontmatter}
\title{Electron reflectivity measurements of Ag adatom concentrations  on W(110)}

\author{J. de la Figuera}
\address{Dpto. de F\'{\i}sica de la Materia Condensada, Universidad Aut\'{o}noma de Madrid, Madrid 28049, Spain}
\ead{juan.delafiguera@uam.es}
\author{N. C. Bartelt,}
\author{K. F. McCarty}
\address{Sandia National Laboratories, Livermore, California 94550, USA}
\date{\today}
\begin{abstract}

The density of  two-dimensional Ag adatom gases on W(110) is determined by monitoring local electron reflectivity using low energy electron microscopy (LEEM). This method of adatom concentration measurement can detect changes in adatom density at least as small as 10$^{-3}$ ML for a $\mu$m size region of the surface. Using this technique at high temperatures, we measure the sublimation rates of Ag adatoms on W(110). At lower temperatures, where Ag adatoms condense into monolayer islands, we determine the temperature dependence of the density of adatoms coexisting with this condensed phase and compare it with previous estimates.
 
\end{abstract}


\end{frontmatter}

\section{Introduction}

The rates of many surface processes -- surface diffusion, for example -- are determined by the concentration of mobile thermal defects. Direct measurements of the concentration of these defects thus are important for our quantitative understanding of surface dynamics. Typically for metal on metal diffusion, the diffusing species are simple adatoms.  In thermal equilibrium, the adatoms are usually present at low densities because their formation energies are on the order of eV's.  The metal adatoms on compact crystal surfaces have high mobilities. The high mobilities make a direct determination of the adatom concentration over a wide temperature range difficult by, for example, proximal probe microscopies such as scanning tunneling microscopy. Only a few methods have been shown to be sensitive to thermal concentrations of adatoms. Two successful macroscopic techniques stand out: work function measurements \cite{KOLACZKIEWICZ1985,KOLACZKIEWICZ1990,Nohlen0}, and reflectivity measurements \cite{Monot1996PRB}. In these methods, the density of adatoms is related to changes in the surface work function (caused by the individual dipoles of the adatoms through the Helmoltz equation) or the specular reflectivity of some probe particle scattered from the surface, respectively. Helium atoms have high cross sections, up to 100 \AA$^2$, to scatter off single adatoms \cite{Poelsema1983,Farias1998} making them a particularly useful probe. A disadvantage of these techniques as applied to date is that they are not spatially resolved, making the measurements sometimes difficult to interpret.

Motivated by this limitation, we have investigated the possibility of using low energy electron microscopy (LEEM) \cite{Bauer1994,Bauer1998,Tromp2000} to measure adatom concentrations. LEEM is a technique in which a beam of low energy electrons is directed towards a surface. A real image of the surface formed by the reflected electrons is magnified with electromagnetic lenses and observed on a detection screen. We use the sensitivity of the electron reflectivity to the adatom density to quantitatively determine the local concentration of Ag adatoms on a W(110) crystal surface. In particular, we employ the imaging capabilities of LEEM to detect the adatom concentration in thermal equilibrium around condensed two-dimensional (2D) Ag islands.

\section{Experimental}

The experiments were carried out in a commercial LEEM system with a base pressure of 7$\times 10^{-11}$ torr. The system has a spatial resolution of 8 nm and uses a heated LaB$_6$ crystal as an electron source. The same surface region can be imaged while heating or cooling.

The W(110) single crystal substrate was cleaned by cycles of exposure to oxygen at 1300~K and brief flashes to 1800~K. Step-free terraces a few microns wide were routinely found on the crystal. Ag was evaporated from a crucible heated by electron bombardment with active temperature control. While dosing at a rate of 150 seconds per ML, the chamber pressure remained below 2$\times 10^{-10}$ torr. The coverage was calibrated by the time required to grow a full layer of the compact phase of Ag on W(110). One monolayer (ML) is defined as the atomic density of this layer \cite{Kim2003} (13$\times 10^{14}$ cm$^{-2}$).

Temperatures above 400$^{\circ}$C were measured using a two-color infrared pyrometer. Below 400$^{\circ}$C, the temperature was measured with a W5\%Re vs W26\%Re (type-C) thermocouple \cite{Smentkowski1996} welded to a washer that was pressed against the sample. Our estimated error in absolute temperature is 10 K. 

The video-rate acquisition system consists of a Peltier-cooled CCD camera whose output is digitized by a commercial digital video system at 30 frames of 720$\times$480 8-bit pixels per second.

\section{Discussion and Results}

\begin{figure}
\centerline{\includegraphics[width=0.8\textwidth]{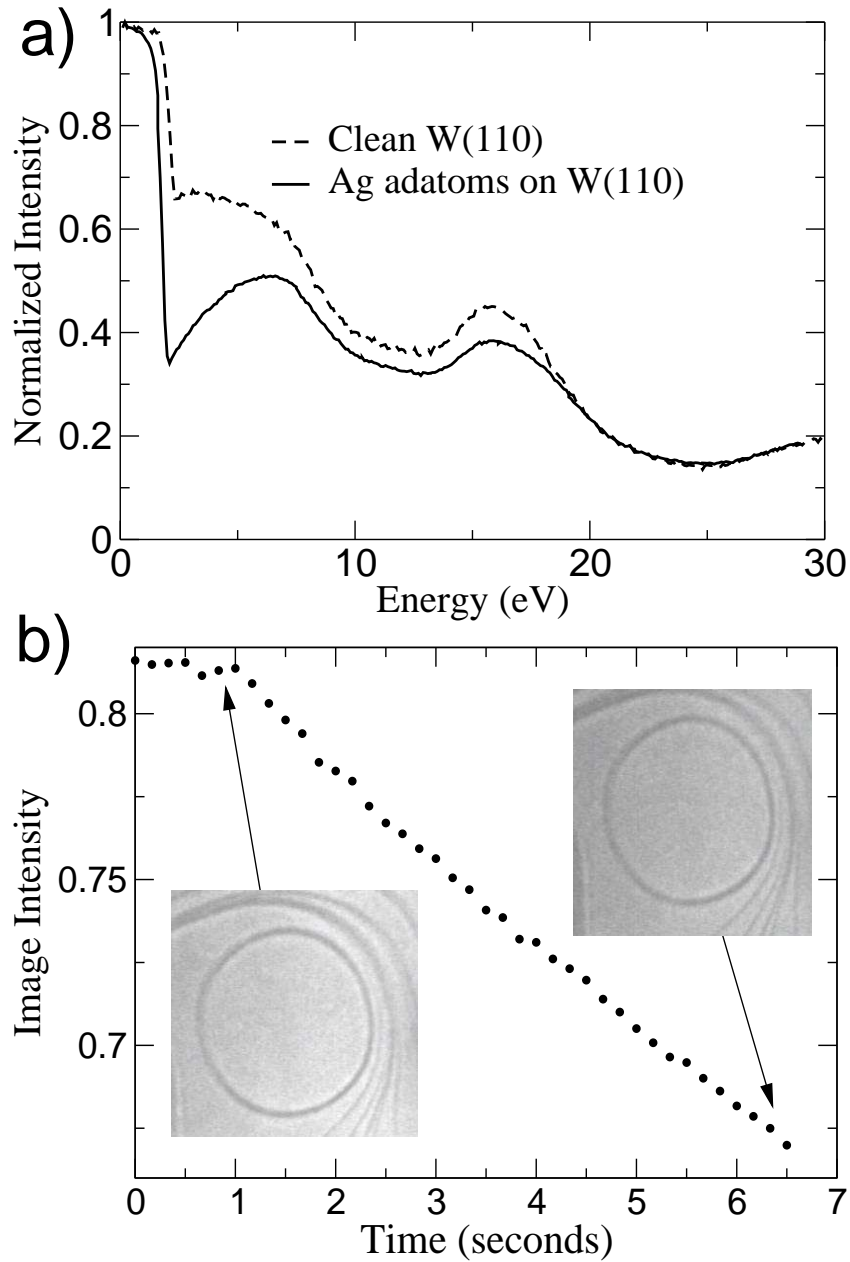}}
\caption{a) Reflected intensity of low energy electrons on clean W(110) (dash line) compared to a W(110) surface covered by a two-dimensional Ag adatom gas (solid line). The intensity is normalized to the intensity at 0~eV. b) Decrease in intensity at 4.9~eV as Ag is deposited on the surface at 670 K, starting at time 1 second. The insets show the LEEM images of the clean W(110) surface and the surface after depositing Ag. Both images are 2.8 $\mu$m wide. The lines in the images correspond to atomic steps and step bunches on the W(110) substrate \cite{Altman1998}. See also Fig.~\ref{growth}.}
\label{reflectivity}
\end{figure}

We first show how the Ag adatoms affect electron reflectivity from W(110). The intensity of the specularly reflected electrons measured from LEEM images for a clean W(110) surface is shown in Fig.~\ref{reflectivity}a for incoming electron energies between 0-30 eV. The intensity variation reproduces previous results going back to the 1960's \cite{Bauer1994}. At energies below the work function, the reflected intensity is high. The peak at 18 eV corresponds to a Bragg reflection. The high intensity between a few eV and about 10 eV likely results from a gap in the projected band structure \cite{Bauer1994}. Ag adatoms decrease the intensity in this energy range.  Figure 1b shows how the reflected intensity changes at a particular energy (4.9~eV, the value used in the remainder of this study) as a function of adatom coverage. An extremely simple dependence is found - the reflected intensity decreases linearly with the Ag adatom concentration. The proportionality between the intensity change and the adatom concentration implies that the effect of each adatom on the reflectivity is independent of the presence of the other adatoms. The linear relationship also allows us to simply determine the adatom concentrations from the electron reflectivity\footnote{The cross section [$(1-\frac{I}{I_o}) = c\times\Sigma$ where $c$ is the adatom concentration and $I$ is the reflected intensity] we obtain for Ag adatoms under the described conditions is $\Sigma = 30 $\AA$^ 2$, similar to the cross section of adatoms in He scattering \cite{Farias1998}.}.

\begin{figure}
\centerline{\includegraphics[width=0.8\textwidth]{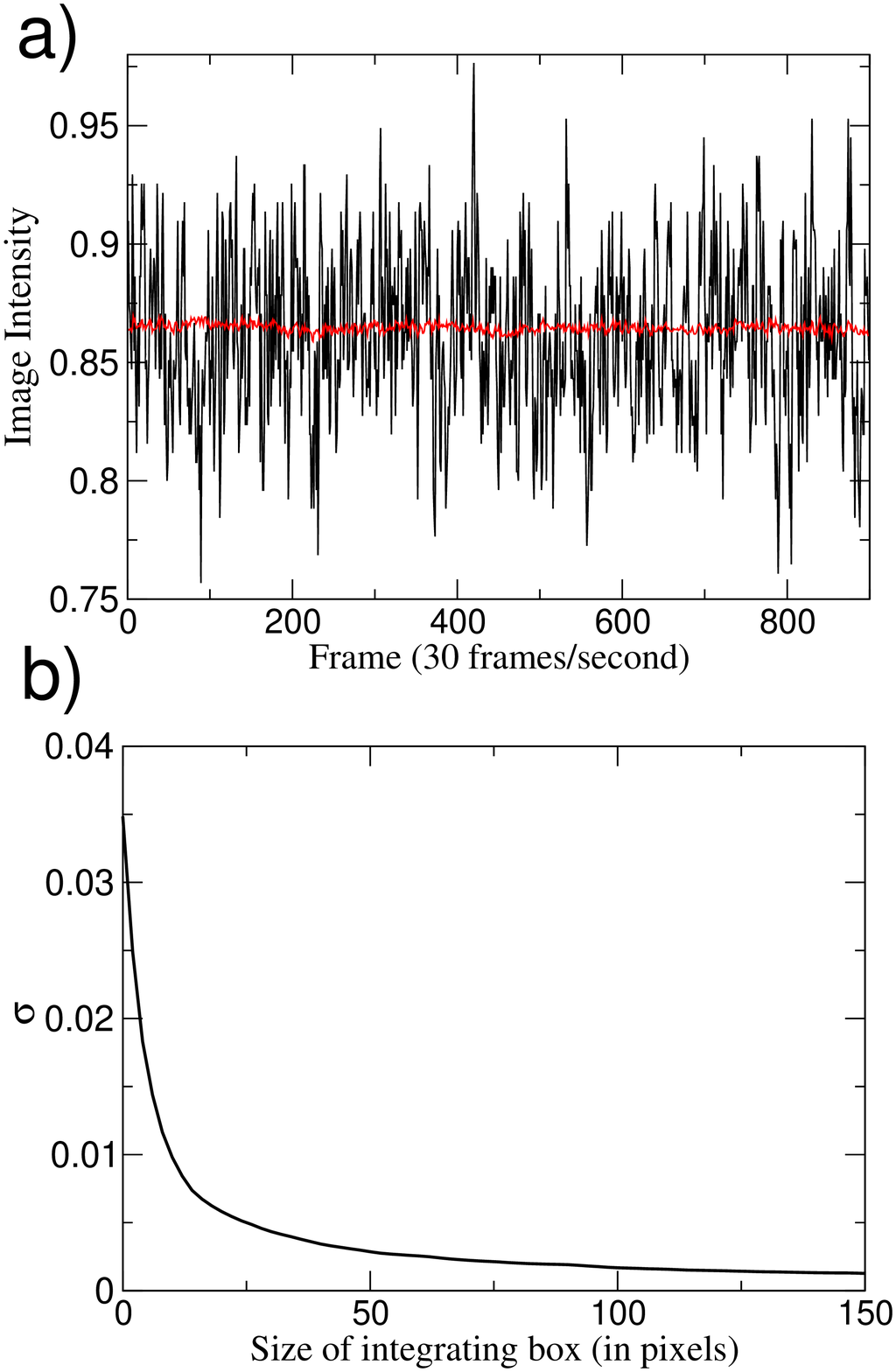}}
\caption{a) Fluctuations in the intensity in a LEEM sequence of images of clean W(110). The solid line corresponds to a single pixel and the red line to integrating the intensity in a square 100 pixel-wide box. Intensity equal to 1 is a saturated 8~bit CCD reading. b) Standard deviation $\sigma$ of the integrated intensity as a function of integrating box width.} 
\label{noise}
\end{figure}

We next estimate how precisely the technique can be used to measure adatom concentration. Under static conditions, the major error source is noise in the LEEM images.  Several sources contribute to the image noise, including noise in: the incident electron current, the channel plates of the detector, and the video detection system used to record the images. Fig.~\ref{noise}a shows consecutive readings at 30 frames per second of the intensity for a single pixel in a 5~$\mu$m LEEM image of clean W(110) (at 4.9~eV). The standard deviation of the measured intensity for the 900 frames shown in the figure is 4\% of the signal. From our intensity calibration, this error translates into an uncertainty in the concentration of Ag adatoms of 2\%. To decrease the effect of this noise, we integrate the intensity over a number of pixels, as shown in Fig.~\ref{noise}b. For example, using a 100 pixel wide box (with a spatial size close to 1$\mu$m) decreases the standard deviation of the intensity to 0.5\% (see Fig.~\ref{noise}). Again, from our calibration, this error corresponds to 2$\times10^{-3}$ ML\footnote{The decrease is smaller than the factor of the square root of the number of pixels, indicating that each pixel is not fluctuating independently.}. Time integration could additionally be used to further reduce the noise and increase the precision. Also we note that the image capture system was in no way optimized for these adatom concentration measurements. However, other sources of errors are present in temperature-dependent experiments: we have observed that changing the temperature by more than a few hundred degrees can produce small changes in the image intensity.  These changes probably arise from changes in the mechanical tilt of the sample. Nevertheless, a precision in the range of 10$^{-3}$ ML with submicron spatial resolution and a 10 ms time resolution can provide useful information,  as we now show. We demonstrate two applications of the method, one of which exploits the acquisition speed, while the other exploits the spatial resolution.

\begin{figure}
\centerline{\includegraphics[width=0.8\textwidth]{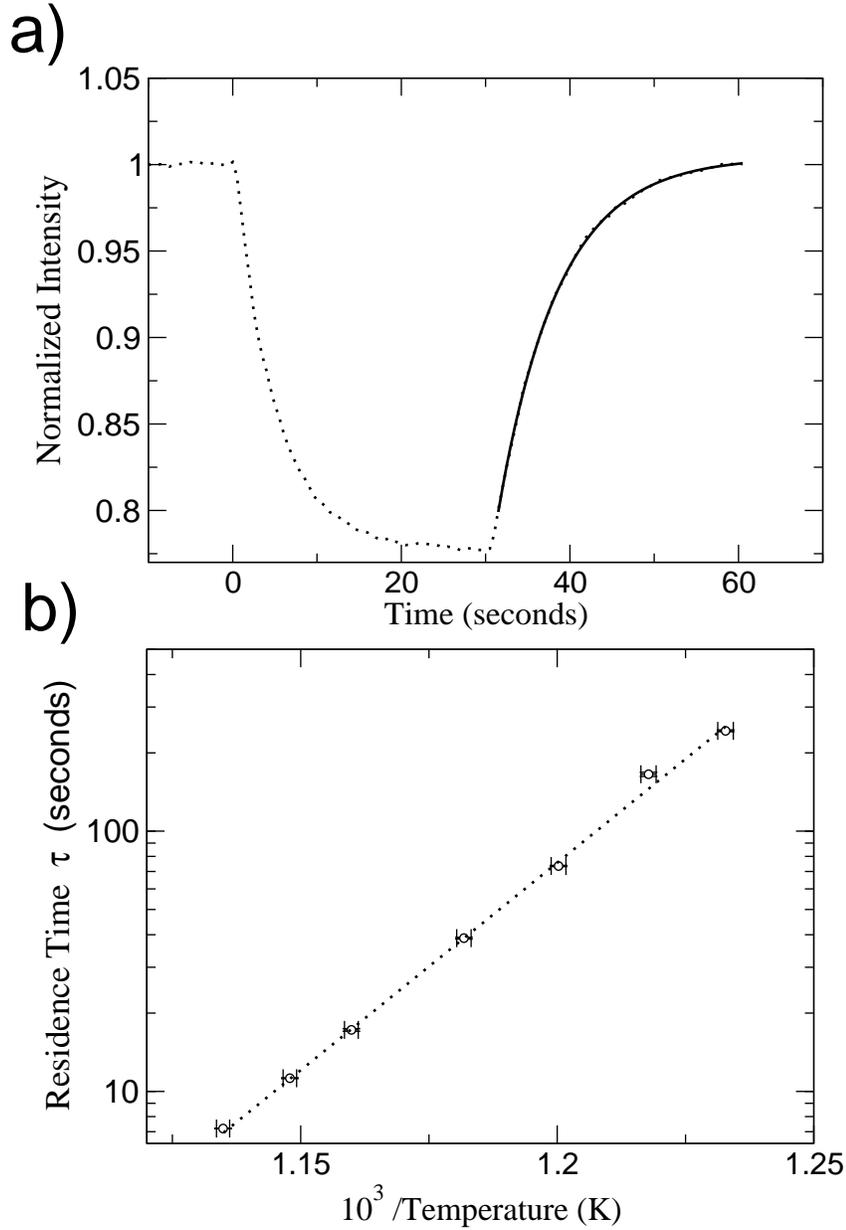}}
\caption{a) evolution of the image intensity (normalized to the starting intensity) while dosing Ag on W(110) at 881 K (dashed line) at 150 seconds/ML between time 0 and time 30 seconds. After the flux is stopped, the intensity recovers as  Ag atoms sublimate. The solid line is a fit to an exponential function from which the residence time of Ag atoms is obtained. b) Plot of the residence times at different temperatures. The line is a fit with slope 3.06$\pm$0.05 eV.} 
\label{decay}
\end{figure}

We first examine the desorption rate of Ag adatoms as a function of temperature. Figure ~\ref{decay}a shows the reflected intensity as W(110) at elevated temperature is exposed to a constant flux of Ag atoms. The intensity initially decreases as adatoms adsorb on the surface. The intensity decay is not linear due to the competition between the incoming flux and desorption from the surface. Eventually, a steady state coverage is reached, which depends on flux and temperature. When the incoming flux is stopped, the intensity gradually recovers its original value as Ag adatoms desorb from the surface. We assume that the concentration of Ag adatoms ($c$) follows $\frac{\partial{c}}{\partial{t}}= -\frac{c}{\tau}$, where $\tau$ is the average residence time of an adatom on the surface. Fitting an exponential function to the experimental intensity curve after the incoming flux has been stopped (Fig.~\ref{decay}a) gives $\tau$ at the given temperature. Repeating the experiment at different temperatures allows the residence time to be measured as a function of temperature. We plot in Fig.~\ref{decay}b the residence time for temperatures in the range between 810 K and 880 K. Because the surface is covered with a single phase (i.e., no condensed Ag is present, unlike the example studied below), we can use a large integration box (100$\times$300 pixels) which gives a noise level in the coverage of only 4$\times10^{-4}$~ML. From the dependence of the residence time with temperature, we estimate the activation barrier for Ag sublimation into the vacuum to be E$_{ev}$= 3.06 $\pm$ 0.05 eV. This number is in agreement with previous estimates by thermal desorption spectroscopy\cite{Bauer1975,Kolaczkiewicz1986}, although we find no need to invoke a coverage dependent activation barrier in our data.

In our next example, we exploit the spatial resolution of LEEM to selectively measure the adatom concentration even when condensed phase Ag exists on the surface. At low coverages and high temperatures, all the Ag adatoms are in a single phase: a 2D adatom gas. At some lower temperature or higher coverage, however, there is a phase transition: Ag atoms condense into monolayer thick islands.  A thermodynamic equilibrium exists between the 2D adatom gas and the coexisting condensed phase. Additional Ag atoms increase the area of the condensed phase. The concentration of the adatom gas in equilibrium with the condensed phase increases with temperature. At a critical temperature, the difference between condensed and adatom phases disappears, and a single fluid phase remains. At sufficiently high temperature, Ag begins to sublime. As shown above, sublimation is negligible at temperatures below 700K.

\begin{figure}
\centerline{\includegraphics[width=0.8\textwidth]{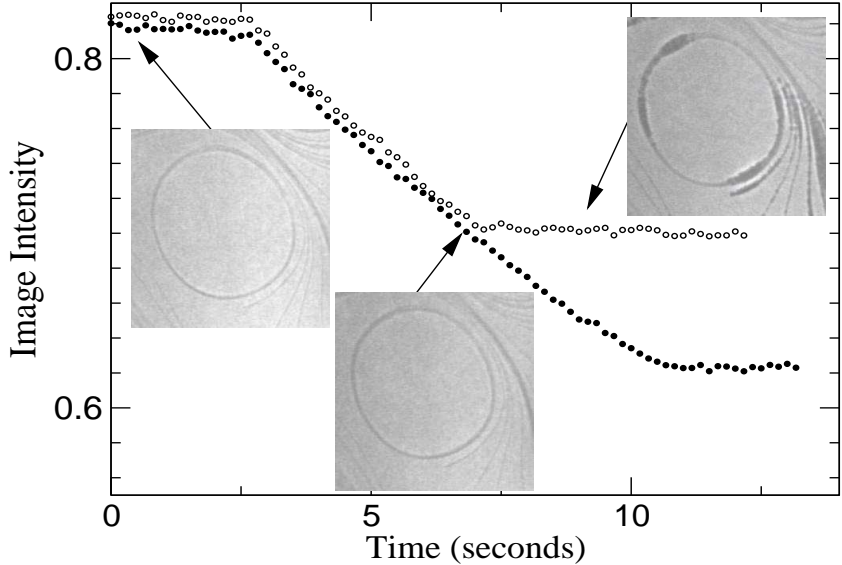}}
\caption{Image intensity on areas not covered by the Ag condensed phase as Ag is deposited. Two different intensity curves are plotted at two different deposition temperatures (filled circles and open circles correspond to  670~K and 600~K, respectively). The insets show three LEEM images selected from the sequence from which the intensity was measured at 600~K. The images are 2.8~$\mu$m wide.}
\label{growth}
\end{figure}

The density of the adatom gas in coexistence with the condensed phase can be determined in the following way\footnote{Because distortions in the reflected intensity occur close (less that 150 nm) to W or Ag steps, the reflectivity measurements are obtained from flat area well separate from steps.}. In Fig.~\ref{growth} we image the surface while depositing Ag. We plot the intensity of regions devoid of condensed Ag.  Initially, the image intensity decreases as the Ag adatom gas concentration increases. As more Ag atoms are deposited, however, the electron intensity abruptly stops changing. Inspection of the LEEM images clearly shows that this happens precisely at the same time that Ag begins to condense on the substrate steps. Subsequent Ag dosing only increases the coverage of the condensed phase.

\begin{figure}
\centerline{\includegraphics[width=0.8\textwidth]{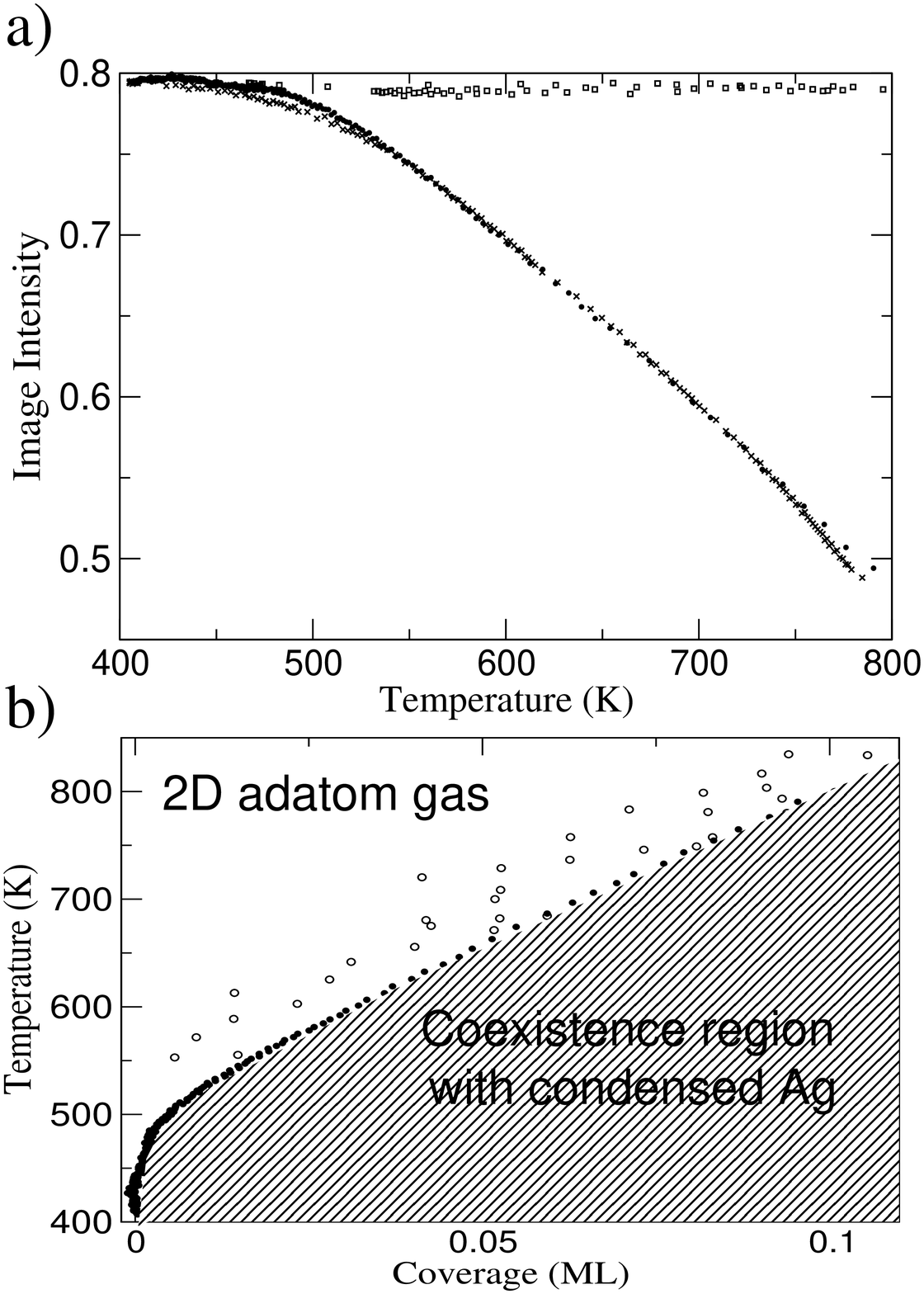}}
\caption{a) Reflected electron intensity from W(110) areas devoid of condensed Ag. during cooling (filled circles) and heating  (crosses). The open squares correspond to the the  reflectivity of clean W(110). b) Coverage at which the Ag adatom gas is in equilibrium with condensed islands of Ag determined by LEEM (filled circles). The open circles correspond to an experimental determination of the equilibrium concentration by means of work function measurements \cite{KOLACZKIEWICZ1985}.}
\label{changetemp}
\end{figure}

Figure 4 also shows that the the condensed phase appears at a easily measured lower coverage at a lower temperature.  Thus, the temperature dependence of the coexistence line between condensed Ag and the Ag adatom gas could be measured by a series of temperature-dependent dosing experiments.  However, there is easier way to determine this temperature dependence. We simply measure the intensity change of the adatom phase as a function of temperature with a total Ag coverage large enough to ensure that condensed phase is present at all temperatures. In Fig.~\ref{changetemp}a the reflected intensity versus temperature is presented.  For comparison, there is nearly no change of reflectivity of clean W(110) with temperature. Using our intensity calibration to obtain the adatom density versus the temperature, we produce the adatom gas-condensed Ag phase diagram. Fig.~\ref{changetemp} shows that there is good agreement between the coexistence lines measured by LEEM and previously by the work function technique \cite{KOLACZKIEWICZ1985}.

\section{Summary}

In this paper we have described a simple method of measuring the temperature dependence of  the concentration of thermal adatom gases using LEEM.   The accuracy of this method will allow the energetics and dynamics of adatoms to be probed in detail.   For example, phase diagrams such as the one shown in Fig.~5 can be compared very closely with statistical mechanical lattice gas models of adsorption to extract information about adatom formation enthalpies and entropies, as well as  adatom-adatom interactions.  In addition to the Ag/W(110) system described in this paper,  we have also successfully applied this technique to measuring thermal Ag and Au adatoms concentrations on Ru (0001) and to Au adatom concentrations on W(110).  A detailed description and interpretation of these observations is in preparation. 

\section*{Acknowledgments}
This research was partly supported by the Office of
Basic Energy Sciences, Division of Materials Sciences,
U. S. Department of Energy under Contract No. DE-AC04-94AL85000, and by
the Spanish Ministry of Science and Technology through Project
No.~MAT2003-08627-C02-02, and by the Comunidad Aut\'onoma de Madrid through Project No.~GR/MAT/0155/2004.


\end{document}